\documentclass[9pt,twocolumn,twoside]{pnas-new}

\templatetype{pnasresearcharticle}

\newcommand{\od}[2]{\frac{\text{d}#1}{\text{d}#2}}

\usepackage{braket}
\usepackage{pdfpages}
\usepackage{placeins}
\begin{document}

\title{Complex multiannual cycles of \textit{Mycoplasma pneumoniae}: persistence and the role of stochasticity}

\author[a,b,c,1]{Bjarke Frost Nielsen}
\author[d,e]{Sang Woo Park}
\author[f]{Emily Howerton}
\author[g]{Olivia Frost Lorentzen}
\author[b]{Mogens H. Jensen}
\author[f]{Bryan T. Grenfell}

\affil[a]{High Meadows Environmental Institute, Princeton University, Princeton, NJ, USA}
\affil[b]{Niels Bohr Institute, University of Copenhagen, Copenhagen, Denmark}
\affil[c]{PandemiX Center, Roskilde University, Denmark}
\affil[d]{Department of Ecology and Evolution, University of Chicago, Chicago, IL, USA}
\affil[e]{School of Biological Sciences, Seoul National University, Seoul, Korea}
\affil[f]{Department of Ecology and Evolutionary Biology, Princeton University, Princeton, NJ, USA}
\affil[g]{Canal Pointe, Princeton, NJ, USA}

\leadauthor{Nielsen}

\significancestatement{The bacterium \textit{Mycoplasma pneumoniae} causes respiratory illness worldwide and is known for its complex and poorly-understood multiannual outbreak cycles. By fitting a mechanistic mathematical model to data from Denmark spanning several decades, we find a parsimonious explanation for the persistent 5-year periodicity, in the form of \textit{quasicycles}. A deterministic model explains the periodicity and shape of the cycles, while environmental randomness -- such as variability in host contact patterns -- is needed to sustain them. The work has practical implications for predicting the epidemiological dynamics of \textit{M. pneumoniae} across diverse settings and suggests that noise-induced dynamics should be considered as a potential driver of complex cycles in other endemic pathogens.
}

\correspondingauthor{\textsuperscript{1}To whom correspondence should be addressed. E-mail: bjarke@princeton.edu}


\begin{abstract} %
The epidemiological dynamics of \textit{Mycoplasma pneumoniae} is characterized by poorly understood complex multiannual cycles. The origins of these cycles have long been debated, and multiple explanations of varying complexity have been suggested. Using Bayesian methods, we fit a dynamical model to half a century of \textit{M. pneumoniae} surveillance data from Denmark (1958–1995, 2010–2025) and uncover a parsimonious explanation for the persistent cycles, based on the theory of quasicycles. The period of the multiannual cycle (approx. 5 years in Denmark) is explained by susceptible replenishment due, primarily, to loss of immunity.  While an excellent fit to shorter time series (a few decades), the deterministic model eventually settles into an annual cycle, unable to reproduce the persistent cycles. We find that environmental stochasticity (e.g., varying contact rates) stabilizes the multiannual cycles and so does demographic noise, at least in smaller or incompletely mixing populations. The temporary disappearance of cycles during 1979--1985 is explained as a consequence of stochastic mode-hopping. The circulation of \textit{M. pneumoniae} was recently disrupted by COVID-19 non-pharmaceutical interventions (NPIs), providing a natural experiment on the effects of large perturbations. Consequently, the effects of NPIs are included in the model and medium-term predictions are explored. Our findings highlight the intrinsic sensitivity of \textit{M. pneumoniae} dynamics to perturbations and interventions, underscoring the limitations for long-term prediction. More generally, our findings provide further evidence for the role of stochasticity as a driver of complex cycles across endemic and recurring pathogens.
\end{abstract}

\dates{This manuscript was compiled on \today}

\maketitle
\thispagestyle{firststyle}
\ifthenelse{\boolean{shortarticle}}{\ifthenelse{\boolean{singlecolumn}}{\abscontentformatted}{\abscontent}}{}

\dropcap{T}he bacterial pathogen \textit{Mycoplasma pneumoniae} is simultaneously common and unusual. It lacks a cell wall and is among the smallest free-living bacteria \cite{kradin2016understandingchap9}. The infection it causes has been given names such as \textit{(primary) atypical pneumonia} \cite{lind1997seroepidemiological}, \textit{cold pneumonia}, and \textit{walking pneumonia} \cite{foy1982mycoplasma} due to its tendency to cause less severe disease compared with typical bacterial pneumonia. It is a very common respiratory pathogen, especially in children and adolescents, accounting for 15--20\% of all cases of pneumonia \cite{foy1993infections}. Disease courses are typically mild, but severe outcomes do occur.  About 5--10\% of upper airway infections with \textit{M. pneumoniae} progress to pneumonia, and there are likely many asymptomatic cases \cite{braun2006mycoplasma}, suggesting a high degree of under-ascertainment.  Extrapulmonary symptoms are increasingly recognized \cite{narita2010pathogenesis,kumar2018mycoplasma} and include complications such as encephalitis and RIME/Stevens-Johnson Syndrome \cite{martinez2024preliminary,canavan2015mycoplasma}, with \textit{M. pneumoniae} being the most common infectious cause of the latter in children \cite{hasbini2024mycoplasma,ravin2007mycoplasma}. As many as 25\% of \textit{M. pneumoniae} cases develop some extrapulmonary symptoms \cite{sanchez2008mycoplasma}, most commonly dermatological or gastrointestinal. Sometimes cardiovascular \cite{asif2023heart} and neurological \cite{cassell1981mycoplasmas} complications are seen, the latter in as many as 1--10\% of those \textit{M. pneumoniae} cases which require hospitalization \cite{waites2004mycoplasma,sanchez2008mycoplasma,bajantri2018mycoplasma}.

From an epidemiological dynamics perspective, a remarkable feature of \textit{M. pneumoniae} are the observed complex outbreak cycles which generally exhibit a 3-7 year period \cite{kim2009mycoplasma,brown2016mycoplasma,wang2017epidemiological,rastawicki1998epidemiology}. In Denmark \cite{lind1997seroepidemiological}, as well as England and Wales \cite{chalker2011increased}, 4--5 year cycles have generally been observed. This is in contrast to some classical, strongly immunizing childhood diseases, such as measles, which tend to exhibit simpler annual or biennial dynamics \cite{grenfell2001travelling}. However, there are exceptions to that rule, for example rubella \cite{rozhnova2013characterizing,keeling2001seasonally}. Despite being such a common pathogen, the mechanisms behind the complex epidemiological dynamics of \textit{M. pneumoniae} remain poorly understood, complicating prediction and forecasting.  
In general, even short and medium term predictability are important for timely resource allocation and deployment of relevant screening and diagnostic testing schemes. However, when the year-to-year incidence exhibits wide variability, planning for outbreaks becomes simultaneously more important and more difficult.  Discerning and understanding patterns in this variability is thus paramount for effective planning, not least for \textit{M. pneumoniae}, where severe outcomes are typically preventable if diagnosed and treated in a timely manner.

Since early 2020, the multiannual cycles of \textit{M. pneumoniae} have been largely disrupted \cite{park2024predicting,sauteur2025global}. Such polymicrobial effects of the COVID-19-related non-pharmaceutical interventions (NPIs) have been observed across many pathogens \cite{chow2023effects}. The most striking example is perhaps the apparent extinction of the Yamagata strain of influenza B virus \cite{koutsakos2021influenza,caini2024probable}.

Over three years, \textit{M. pneumoniae} circulation was drastically curtailed in many locations, and Denmark is no exception. A similar multi-year delay in \textit{M. pneumoniae} resurgence has recently been reported for Europe and Asia \cite{sauteur2024mycoplasma} as well as the United States \cite{park2024predicting,boyanton2024sars}. Such large-scale perturbations effectively provide a ``natural experiment'', demonstrating how the dynamics of \textit{M. pneumoniae} responds to a prolonged disruption and the associated build-up of susceptible individuals. A recent study of respiratory syncytial virus (RSV) and human metapneumovirus (hMPV) found that including the NPIs the and post-NPI rebound could aid model parameter estimation \cite{howerton2025using}, since including data from outside the regular, cyclic regime tends to improve identifiability.

While the origins of the multiannual cycles of \textit{M. pneumoniae} are generally regarded as unclear \cite{xu2024molecular,chalker2011increased}, a number of different mechanisms of varying complexity have been proposed in the literature. Among the earlier ones was the suggestion by \cite{lind1997seroepidemiological} that immunity following infection simply has approximately the same duration as the observed cycle. This explanation, however, has difficulty accounting for how cycle periods can differ somewhat between countries. Without accounting for susceptible dynamics in detail, it is also unclear why the immunity duration should have to match the cycle period. Zhang et al. \cite{zhang2019positively} suggested a combination of strain dynamics, contact network structure and waning immunity as the explanation, while Omori et al. \cite{omori2015determinant} explored three different possible explanations: school-term forcing, a low-dispersion distribution of immunity duration, and interference between \textit{M. pneumoniae} serotypes. They find that, of these three explanations, a less variable duration of immunity (compared with the typically assumed exponential waning) best captures the generally observed patterns in that it allows for incidence oscillations.   
Recently, the authors of \cite{sauteur2025global} used a TSIR model to describe the post-COVID rebound of \textit{M. pneumoniae} in Denmark as well as four UN regions (Europe, Asia, the Americas, and Oceania) but found that they required at least a 90\% reduction in the transmission rate due to non-pharmaceutical interventions in order to fit the data.

Here, we use a mathematical model that incorporates seasonality, vital dynamics (births/deaths), immunity loss and the effects of NPIs to study recent (2010–2025) and historical (1958–1995) \textit{M. pneumoniae} incidence data from Denmark. The simple deterministic model is found to capture the observed dynamics well over a few decades but eventually reverts to annual outbreaks, failing to reproduce the prominent multiannual cycles long-term. We show that the presence of environmental noise can indefinitely sustain the multi-year outbreak patterns. In doing so, we highlight the important role that fluctuations in contact patterns, environmental noise, and demographic processes play when forecasting pathogens with complex dynamics such as \textit{M. pneumoniae}.

\section*{Lessons from contemporary and historical data}
We utilize two different data sets as correlates of \textit{M. pneumoniae} incidence, both from Denmark (Statens Serum Institut, SSI). One has \textit{M. pneumoniae}-positive PCR tests (and the raw number of tests performed) with a weekly reporting frequency and covers the period 2010--2025. This data set forms the primary basis for parameter estimation, since the test data is highly \textit{M. pneumoniae}-specific and temporally well-resolved. Secondarily, we use a data set covering the period 1958--1995. Beginning in 1946, patient blood samples collected at hospitals and general practitioners for the diagnosis of Primary Atypical Pneumonia (later identified as \textit{M. pneumoniae}) were sent to SSI to undergo the cold agglutinin (CA) reaction. This reaction, unfortunately, has only low to moderate sensitivity ($\sim 20-80\%$ \cite{fischer2018cold, lind1997seroepidemiological}) and moderate specificity ($\sim 82-92\%$ \cite{fischer2018cold, lind1997seroepidemiological}) for \textit{M. pneumoniae}. However, beginning in 1958, CA-positive samples were also subjected to \textit{M. pneumoniae} Complement Fixation (MpCF), which was found to have higher specificity \cite{grayston1965mycoplasma,beersma2005evaluation}. We thus also treat those tests from the 1958--1995 data set which are \textit{simultaneously} CA- and MpCF-positive as an incidence proxy on which we base (secondary) parameter estimates.  

The duration of immunity following \textit{M. pneumoniae} infection is known not to be life-long, but the precise duration is unknown, with estimates ranging from a few years to more than a decade \cite{foy1983naturally,lind1997seroepidemiological,park2024predicting}. For this reason, we opted for an SIRS (\textbf{S}usceptible-\textbf{I}nfected-\textbf{R}ecovered-\textbf{S}usceptible) compartmental model which allows previously infected individuals to become susceptible, and thus potentially reinfected. The model includes seasonal modulation of the transmission rate and equal birth/death rates (which are assumed constant within each data fit, for simplicity). The duration of immunity was treated as a parameter to be estimated during fitting to data.

The influence of NPIs is implemented as a slowly-varying multiplicative effect on the transmission rate from March 2020 onward, representing a reduction in overall contact rates, the magnitude of which is an estimated parameter (see Supplementary Fig. S7). %

We estimate a basic reproductive number of $R_0=1.6$ (95\%CI [1.45, 1.82]), in good agreement with ref. \cite{zhao2019phase}, a mean duration of immunity of 8--9 years and a relative amplitude of seasonality in transmission of $0.18$. All estimated parameters and their uncertainties can be found in Table \ref{tab:params}.

With a sinusoidal seasonality function\footnote{We opted for a sinusoidal seasonality function for simplicity, but have also directly estimated the shape of the seasonality function from the data. The resulting seasonality function can be found in Supplementary Fig. S5.}, %
the model fits the 2010--2025 data well (with the partial exception of the 2017 season in which the apparent incidence was somewhat higher than modeled), see Fig. \ref{fig:fit}A. The historical 1958--1995 data, which is overall of much lower fidelity, is captured more loosely by the model, see Fig. \ref{fig:fit}B. Interestingly, here we had to allow for a slow change in the mean transmission rate and ascertainment (allowed to change only slightly every 4 years), otherwise the model would tend to settle into an annual pattern.

This leads us to a more general observation: the SIRS model with \textit{M. pneumoniae} parameters is able to capture the patterns reasonably well over a few decades, but in the absence of perturbations, the modeled dynamics always decays into an annual pattern eventually. This is of course at odds with the ubiquity of complex multiannual cycles in \textit{M. pneumoniae}  \cite{lind1997seroepidemiological,kim2009mycoplasma,brown2016mycoplasma,wang2017epidemiological,rastawicki1998epidemiology,chalker2011increased} and other SIRS-like epidemiological systems such as parvovirus B19 \cite{russcher2024changing,nicolay2009clinical,bosman2002fifth,enders2007current,vyse2007burden}. In the next section we discuss how the time-scales of the \textit{M. pneumoniae} system may give rise to an approximately 5-year cycle, at least transiently. We then go on to investigate the role that stochasticity plays in sustaining such cycles over decades and in re-igniting them even after periods of approximately annual dynamics. In the final part of the paper, we fit a stochastic model to the data and study projections of future \textit{M. pneumoniae} incidence with and without stochasticity.

\subsection*{The origins of intrinsic cycles}
The seasonally forced SIRS system with vital dynamics contains two comparable oscillatory time-scales: the one-year seasonal period and the time-scale set by the intrinsic epidemiological dynamics. To ascertain the latter time-scale, the corresponding autonomous system may be studied (that is, the SIRS system with vital dynamics but no seasonal forcing). This system will eventually reach an endemic equilibrium, but will perform underdamped oscillations as it approaches this equilibrium. 
A linearization of the autonomous system around its endemic equilibrium yields an intrinsic oscillation frequency as well as an exponential decay rate \cite{keeling2008modeling}, given by a combination of the transmission rate $\beta$, immune waning rate $\delta$, recovery rate $\gamma$ and birth/death rate $\mu$. See Eq. \ref{eq:Tintrinsic} in \textit{Methods and Materials} for the analytic expression for the intrinsic period.
Using the parameter values estimated from the Bayesian fit to Danish data (2010--2025), the intrinsic period is found to be $T = 5.0\text{yr}$ (95\%CI: [4.96, 5.10]).
In Supplementary Fig. S6, %
we explore the pairwise parameter dependence of the intrinsic period. The parameter values estimated by fitting to the 2010--2025 data are indicated with black crosses. The intrinsic period depends strongly on the transmission rate ($\beta$), infectious period ($1/\gamma$) and immunity duration ($1/\delta$), while the birth/death rate ($\mu$) plays only a minor role. Since the mean duration of immunity and infectious period are primarily a function of intrinsic host-pathogen interactions, we conclude that the primary determinant of periodicity in different countries is likely to be the transmission rate.

Grossman remarked \cite{grossman1980oscillatory,keeling2008modeling} that the SIR equilibrium (w. vital dynamics) tends to be only weakly stable, since the real part of the eigenvalue (the decay rate) is much smaller than the imaginary part (the frequency). We find the same phenomenon for the SIRS system with \textit{M. pneumoniae} parameters, with the ratio $\vert\text{Im}\{\lambda\}/\text{Re}\{\lambda\}\vert \approx 12$ (with $\lambda$ being the relevant eigenvalue of the Jacobian matrix for the linearized system).

A simple Fourier transform of the 2010--2020 (pre-COVID) data yields a dominant period of 5.1 years, while a Lomb-Scargle (LS) periodogram \cite{robitaille2013astropy}, which has finer frequency resolution, places the peak at 4.9 years. For the MpCF-based 1958--1995 time series, the LS periodogram shows the annual signal as being the strongest, which makes sense given the observation of \cite{lind1997seroepidemiological} that the multiannual outbreak pattern was somewhat overtaken by annual dynamics during the late 1970's and early 1980's. However, the strongest multiannual peak for the whole period (1958--1995) is at 4.1yrs (LS)/4.2yrs (Fourier). If the 1958--1975 period is viewed in isolation, the strongest peak (eclipsing the annual signal as well) is at 4.6yrs (LS)/4.3yrs (Fourier). See Supplementary Fig. S2 %
for Fourier and LS spectra of the incidence time series, as well as a wavelet analysis which shows changes in the dominant frequency over time. 

In the next two sections, we consider the role that stochasticity may play in sustaining the observed multiannual cycles. Apart from stochasticity being a fact of nature, the introduction of a stochastic model is further motivated by the observation that the fit in Fig. \ref{fig:fit}A displays low uncertainty compared with the variability in the observed data, suggesting that noise is needed to explain the data.

\section*{Demographic stochasticity}
Spurred by the observation that the deterministic model will always eventually decay into annual dynamics, it is natural to ask how a departure from determinism affects the dynamics.
Demographic stochasticity is a ubiquitous source of noise in population dynamics (including epidemiological dynamics) \cite{engen1998demographic,grenfell2002dynamics}. This type of stochasticity occurs in finite populations and is fundamentally due to two factors. 1) Individuals being discrete (as opposed to the real numbers $S$, $I$ and $R$ of a typical differential equation-based model), and 2) The presence of probability \textit{rates} such as $\mu$, $\beta$, $\gamma$ and $\delta$, which implies stochasticity in the timing of birth/death, infection, recovery and immune waning in finite populations. \\
We implement demographic stochasticity by considering an SIRS model (with vital dynamics) in which individuals are discrete, and may transition between compartments at rates given by \eqref{eq:gilrates} (\textit{Materials and Methods}). Following the approach of \cite{alonso2002extinction,alonso2007stochastic,wang2012simple}, based on the van Kampen system size expansion, we derive an expression for the power spectral density (PSD) $P(\omega)$ of noise-induced fluctuations in the (non-seasonal) SIRS model with demographic stochasticity:
\begin{align}\label{eq:psd}
    P(\omega) = \frac{\alpha + B \omega^2}{(\omega^2-\Omega_0^2)^2 + \Gamma^2 \omega^2}.
\end{align}
Here, $\alpha$, $B$, $\Omega_0^2$, and $\Gamma^2$ are all functions of the parameters $\beta$, $\gamma$, $\delta$ and $\mu$. Their expressions are given in the supplement along with a derivation of \eqref{eq:psd}. In \ref{fig:demographic}A, we plot the PSD with parameters from the 2010--2025 fit, and note that while the peak is at five years, a range of frequencies are excited by demographic noise.

By simulating the seasonally-forced stochastic SIRS model using the Doob-Gillespie algorithm (details in \textit{Materials and Methods}) \cite{gillespie1977exact}, we find that multi-year cycles may indeed be sustained by demographic noise in sufficiently small populations, suggesting that the system exhibits a stochastic resonance \cite{keeling2008modeling}. However, the effects of demographic stochasticity are weaker in larger populations, and clear multiannual cycles can only be sustained by this mechanism in populations of less than approximately 2 million individuals, see Fig. \ref{fig:demographic}. Simple demographic stochasticity alone is thus unlikely to be the source of sustained multiannual cycles in \textit{M. pneumoniae}, at least as far as well-mixed populations are considered. In the supplement, we consider a metapopulation version of the SIRS model with demographic stochasticity and show that multiannual cycles may persist in larger, incompletely mixing populations (see Supplementary Fig. S11). %

\section*{Environmental stochasticity}
We go on to consider the effects of \textit{extra-demographic} or \textit{environmental} stochasticity. Random fluctuations in the transmission rate $\beta(t)$ will be the primary source of environmental stochasticity considered, but we will also explore the effects of more generic exogenous shocks to the system. Such perturbations can have many different origins, including changes in contact and mobility patterns (such as during holiday periods or due to large events), superspreading of biological and behavioral origin \cite{nielsen2023counterintuitive,althouse2020superspreading,sneppen2021overdispersion,nielsen2021covid}, as well as phenotypic variation in the pathogen itself due to mutations.
We implement a version of the seasonal SIRS model with stochastic transmission rates based on the method given in \cite{he2010plug} (see \textit{Materials and Methods}). This SIRS model with environmental stochasticity is then used for parameter estimation as well as direct simulation. The parameters obtained by fitting this model to 2010--2025 data can be found in Table \ref{tab:params}, labeled ``stoch.''.

As shown in Fig. \ref{fig:betafluc}, random fluctuations in the transmission rate tend to stimulate approximately 5-year cycles for a wide range of moderate noise levels. When stochasticity is very slight, the annual cycle dominates (e.g. $\sigma=0.05$ in Fig. \ref{fig:betafluc}), and when environmental fluctuations are very strong, the epidemic dynamics lacks predictable patterns and becomes dominated by large but rare outbreaks, appearing as a Fourier spectrum with all frequencies represented and no discernible peaks (see e.g. $\sigma=0.5$ in Fig. \ref{fig:betafluc}).
It should be noted that even a subdominant multiannual peak in the Fourier spectrum (with e.g. half of the peak power of the annual signal) leads to a prevalence time series with a clear multiannual component. The averaging process of Fig. \ref{fig:betafluc} leads to broader and flatter multiannual peaks than those found in individual spectra, but clearly shows the range of excited periodicities.

While the 5-year cycle is the most typical multiannual cycle at moderate noise levels (given \textit{M. pneumoniae} parameters), other cycles are excited as well, as shown in Fig. \ref{fig:betafluc}, with approximately three-year as well as 6--7 year cycles being somewhat common at moderate-to-high noise levels ($\sigma > 0.2$). In Supplementary Fig. S4, %
we show how the addition of continuous environmental noise to an initially purely annual system can rapidly excite the multiannual cycles.

The immediate impact of a singular perturbation (a ``kick'') to the system \cite{rohani2002interplay} is explored in Fig. \ref{fig:kicktiming}. The perturbation is implemented as either a discrete change in the infected population (Fig. \ref{fig:kicktiming}A-C) or the susceptible population (Fig. \ref{fig:kicktiming}D). Since just a single ``kick'' is  given, the system will only transiently settle into a multiannual cycle and we thus measure the strength of the 5-year cycle during the 20 years immediately following the perturbation.  In Fig. \ref{fig:kicktiming}A, we show the trajectory of the system in the $(S, I)$ configuration space immediately before and after a single kick to the system. In this case, the perturbation consists of a sudden reduction in prevalence $I(t)$, but a sudden change in the susceptible population can have a similar effect (Fig. \ref{fig:kicktiming}D). The annual attractor, where the system resides prior to the perturbation, appears as the central ellipse, while the resulting (transient) five-year cycle encloses it. The rates at which $I(t)$ and $S(t)$ change is reflected in the spacing between the colored markers which are placed one week apart. The colors of the markers themselves indicate the time of year. As is evident, the five-year cycle can be quite complex. The perturbation that initiated the five-year cycle is visible as the approximately vertical line connecting the annual and multiannual cycles, and the state of the system immediately prior to the perturbation is indicated by the red cross. 
Fig. \ref{fig:kicktiming}B shows that the timing of the perturbation matters, with a kick during the low-transmission part of the season most likely to excite the multiannual cycle.  In other words, a kick of a certain size (whether in the positive or negative direction) has a larger effect if it occurs when circulation is already low. 
However, Fig. \ref{fig:kicktiming}C reveals that this seasonal dependence is completely accounted for by the \textit{relative} magnitude of the perturbation, $\Delta I/I$, i.e. the change in incidence divided by the current incidence at that time. If plotted as a function of this relative magnitude, the strength of the excited cycle shows only a very faint seasonal dependence. Fig. \ref{fig:kicktiming}D is analogous to \ref{fig:kicktiming}B, except that the perturbation is applied to the susceptible fraction $S(t)$. Here, seasonal dependence is fairly minimal, since the susceptible fraction fluctuates much less during a season (relative to its mean value) than the prevalence $I(t)$ does.

The mechanism of sustained cycling demonstrated here is fundamentally that of a quasicycle, a mechanism of population cycling proposed in \cite{nisbet1976simple}. It arises from an otherwise stable, underdamped system being subjected to noise which excites an intrinsic cycle that would otherwise decay. In other words, noise continually disrupts the system's deterministic approach towards equilibrium.  In Supplementary Fig. S8, %
we demonstrate that the same phenomenon occurs in the absence of seasonal forcing.
In the ecological literature, there is a fundamental distinction between phase-remembering and phase-forgetting quasicycles \cite{nisbet_gurney_1982,pascual2003quasicycles,turchin1992complex}.
In the former case, the cycle retains phase information over time, such that even temporally well-separated fluctuations exhibit a substantial correlation, typically due to an external periodic forcing. In the phase-forgetting case, this long-term synchronization is eventually lost. In Supplementary Fig. S9, %
we measure the autocorrelation of the prevalence signal $I(t)$ in the fitted model and find that the multiannual cycles are of the phase-forgetting kind. If one considers only the autocorrelation of the raw prevalence time series (Fig. S9A), %
one could be led to believe that the system is phase-remembering, since a substantial autocorrelation remains at large lags $\tau$. However, as we show in Fig. S9B, %
this residual autocorrelation is solely due to the annual modulation, and any phase information in the multiannual cycle is lost over time. We find that the autocorrelation of the multiannual cycle decays approximately exponentially with a half-life of 6-7 years.

The finding that environmental stochasticity can sustain the multiannual cycles of the system is significant because, unlike demographic stochasticity, this mechanism is independent of population sizes. 

Aside from explaining the existence of multiannual cycles, the stochastic model also explains the temporary disappearance of the multiannual cycles between the late 1970s and mid-1980s, observed by \cite{lind1997seroepidemiological}. As shown in the wavelet analysis of model results in Supplementary Fig. S10, %
the stochastic model exhibits mode-hopping, i.e. spontaneous and temporary reversions to the annual attractor, similar to what can be seen in the wavelet transformed empirical data of Supplementary Fig. S2. %
The model fit of Fig. \ref{fig:fit}B also partially captures this phenomenon, with small infrequent changes in the transmission rate in lieu of continuous noise.

\section*{Beyond the post-COVID rebound of \textit{M. pneumoniae}}
Although the deterministic seasonal SIRS model captures the 2010--2025 dynamics well, predicting or forecasting \textit{M. pneumoniae} incidence may be challenging. We know that long-term predictions using the deterministic model are limited by its tendency to revert to annual dynamics. We have also established that phase information in the multiannual cycles is lost over time when the system is subject to stochasticity, which the real-world system invariably is. Both of these findings suggest that a deterministic model is fundamentally limited in terms of long-term predictive power. In the short term, however, a deterministic model may still be appropriate for forecasting.
Fig. \ref{fig:prediction}A shows the trajectory obtained by continuing the deterministic seasonal SIRS model for 10 years beyond the 2010--2025 data (with parameters fitted to said data). 
However, the low uncertainty of the prediction offered by the deterministic model may be an underestimation. Fig. \ref{fig:prediction}B shows the fit of the seasonal SIRS model with environmental stochasticity. While the 2010--2025 data is still captured well, prediction beyond a year or two is associated with high uncertainty. The noise level was estimated at $\sigma=0.12$ using this model, a degree of stochasticity capable of sustaining pronounced multiannual cycles, as we saw in Fig. \ref{fig:betafluc}. We note that this level of stochasticity is substantial, in that $\sigma$ is comparable in magnitude to the estimated amplitude $a$ of seasonality at approx. $0.18$.

The sensitivity to noise can be measured via the Lyapunov exponents of the system. A Lyapunov exponent $\lambda$ measures sensitivity to initial conditions -- this also includes the tendency of the system to exponentially amplify ($\lambda>0$) or buffer ($\lambda<0$) perturbations/noise \cite{pascual2003quasicycles}. Supplementary Fig. S1 %
explores the local/instantaneous Lyapunov exponents (LLEs) \cite{jensen1991intermittency} of the deterministic as well as stochastic model and shows that, while perturbations tend to be transiently amplified, all LLEs eventually become negative, indicating the eventual decay of the perturbation.

While sensitivity to stochastic effects may complicate medium-term predictions (note the spread in predictions in \ref{fig:prediction}B from 2027 onward), some short-term features are relatively robust even at this noise level. Predictions agree that the 2025-2026 season is likely to be a moderate-to-high incidence year, exemplifying how deterministic models yield valuable short-term forecasts even in noise-driven epidemiological systems.

While long-term predictions using the deterministic model are limited by its decay into annual dynamics, the presence of stochasticity limits extended prediction in general, not only by causing variability in the incidence during any given year, but also by enabling mode-hopping, i.e. transitions between annual and multiannual cycles \cite{heltberg2016noise,grenfell2002dynamics}, which appear in the empirical data (Supplementary Fig. S2) %
as well as in model simulations (Supplementary Fig. S10). %
That the predictions for the years immediately following the large NPI-induced perturbation agree well between the stochastic and deterministic models fits well with the general observation that restorative forces are stronger far from equilibrium, leading the stochastic model to (temporarily) behave more like the deterministic model \cite{keeling2008modeling}.

\section*{Limitations}
In fitting the 2010--2025 time series, we used the PCR positivity rate as an incidence proxy, while we used the number of positive tests for the 1958--1995 data (total number of tests were not on record). Neither of these measures are perfect incidence proxies, and they suffer in distinct ways from biases introduced by varying testing intensities. If testing intensity is increased, this leads to an increase in the \textit{number} of positive tests, while it leads to a decrease in the test positivity \textit{rate} (since the denominator -- the total number of tests -- is increased). It may be possible to enhance fitting further by accounting for fluctuations in testing intensity when such data is available, something we have not attempted to do, beyond simple variance stabilization. Furthermore, the fidelity of the 1958--1995 data series is affected by both the (somewhat imperfect) test characteristics of the cold agglutination and complement fixation tests (since the samples considered positive are those which were positive in \textit{both} of these tests \cite{lind1997seroepidemiological}). We have not taken age structure into account \cite{nordholm2024mycoplasma}, nor have we considered detailed spatial effects which might enhance the impact of demographic stochasticity \cite{ovaskainen2006space}.

\section*{Discussion and conclusion}
The origins of non-trivial cycles in ecological systems and the extent to which stochasticity plays a role in shaping them are long-standing questions in ecology, including epidemiology \cite{rohani2002interplay,coulson2001age,grenfell1998noise}. While the concept of quasicycles -- periodic population fluctuations resulting from the interaction of noise with decaying intrinsic cycles -- was proposed by \cite{nisbet1976population}, the theory has evolved since. Significant analytic advances for SIR-type systems \cite{alonso2007stochastic} and predator-prey systems \cite{pascual2003quasicycles,mckane2005predator} were made two decades ago, and quasicycles were proposed as a model of seasonal fluctuations in influenza incidence around the same time \cite{dushoff2004dynamical}, albeit without empirical verification.
In \cite{wang2012simple}, demographic stochasticity was considered as a potential driver of recurrent outbreaks of avian influenza in North American wild waterfowl and the observed cycles were quantified through wavelet analysis. The authors used a stochastic SIR-type model, extended to include environmental transmission (but not environmental stochasticity), and found that it could explain the occurrence of multiannual outbreak cycles, with the period being largely determined by the intensity of environmental transmission.

The effects of demographic stochasticity have been most widely considered, while environmental noise features less prominently in the epidemiological literature, and studies that consider both are rare. Pertussis (whooping cough) is another example of a disease where multiannual cycles have been studied in detail. In \cite{broutin2005large}, the authors used wavelet analysis to study the multiannual cycles of pertussis across 12 countries and noted that, while a 3--4 year periodicity was typically found (a result echoed by \cite{metcalf2009seasonality}), it is transient in nature. The multiannual cycles of pertussis were previously shown to likely originate from the system's response to dynamical noise \cite{rohani1999opposite}, and a study found that stochasticity is essential to even qualitatively capture the cycles of pertussis \cite{rohani2002interplay}. 

For \textit{M. pneumoniae}, we find that noise-perturbed intrinsic dynamics is sufficient to explain the complex cycles, offering a parsimonious model for the dynamics. Importantly, the model also explains the temporary disappearance of the multiannual cycles observed in the Danish data from the late 1970s until the mid-1980s as a consequence of mode-hopping, i.e. the system reverting to the annual attractor due to a perturbation. This temporary disappearance of the multiannual pattern was previously noted by \cite{lind1997seroepidemiological}, who proposed changes in daycare attendance as a possible driver. 

Earn et al. \cite{earn2000simple} demonstrated the existence of a rich landscape of stable cycles in the related, seasonally driven SEIR model at high transmission rates (basic reproductive numbers in excess of approx. 4). In the current case of \textit{M. pneumoniae} however, multiannual cycles are unstable and require stochasticity or perturbations to be sustained. A similar systematic mapping of possible cycles in the seasonal SIRS model -- taking stochasticity into account and including subharmonic resonances \cite{nisbet1976population,guckenheimer2013nonlinear} -- would be highly useful.

With the COVID-19 NPIs phased out, another major -- if less abrupt -- change in \textit{M. pneumoniae} epidemiology is on the horizon. Macrolide resistance has been growing rapidly in recent years, albeit with large geographical differences \cite{kim2022global}. It is currently unclear how this will affect epidemiological dynamics, but some studies have found macrolide-resistant strains to be associated with increased virulence and longer disease duration \cite{zhou2014more,lee2023comparison}, leading us to speculate that average transmissibility could be influenced as well. Extending the current model to incorporate observational data on macrolide resistance would be highly worthwhile, and holds the potential to estimate any effects of macrolide-resistant strains on \textit{M. pneumoniae} dynamics.

Since the main objective of this study was to determine the mechanism of population cycling in \textit{M. pneumoniae}, we opted for a fully mechanistic model. This has the further advantage of enabling straight-forward prediction of the overall dynamics in parameter ranges that differ from those obtained from the Danish data. Some parameters are likely intrinsic to the pathogen-host interaction (rates of waning and recovery), while others (transmission rates and birth/death rates) are likely to vary geographically, leading to different dominant periods according to Eq. \ref{eq:Tintrinsic}. For adaptive forecasting, however, the formulation of a hybrid semi-mechanistic machine learning model would likely be advantageous. In a recent study \cite{lau2022comparing}, a machine learning model was shown to perform well when tasked with short- and medium-term forecasting of historical measles data. The authors note that {``[t]he primary focus of mechanistic models has been \textit{understanding and characterizing} the natural history of transmission [\dots], statistical and machine learning techniques in infectious disease modeling have primarily focused on improving \textit{forecasting} accuracy without the explicit aim of inferring transmission dynamics.''} While beyond the scope of this study, we encourage a synthesis of the two approaches for \textit{M. pneumoniae} forecasting.

In this study, we have considered both demographic and environmental stochasticity as potential drivers of multiannual outbreak cycles. However, we have not attempted any detailed attribution of stochastic effects to behavioral or demographic sources. A future study could conceivably pinpoint the sources of stochasticity that most strongly shape \textit{M. pneumoniae} dynamics by extending the current work to incorporate (time-varying) population structure (in terms of age and contact patterns) and mobility data.

In summary, we have offered a parsimonious explanation for the multiannual cycles of \textit{M. pneumoniae} as a type of quasicycle, and provided a computational and analytic framework for similar analyses of other pathogens and locations. There is a need for comparative studies of the factors governing the diversity of complex cycles observed globally for pathogens such as \textit{M. pneumoniae} and parvovirus B19, across countries and pathogens. Such comparisons may also enable co-estimation of environmental sources of stochasticity, and the framework presented here is well suited to facilitate this.

\begin{table}%
\centering
\caption{Estimated and assumed parameters from fit to 2010--2025 \textit{Mycoplasma pneumoniae} surveillance data from Denmark. parameters labeled ``determ.'' are estimated using the deterministic seasonal SIRS model, while those labeled ``stoch.'' are estimated using the seasonal SIRS model with stochastic transmission rate.  }
\label{tab:params}
\begin{tabular}{lrr}
Parameter & Description & Estimated value [95\% HPD]\\
\midrule
$\beta_0$ (determ.) & Mean transmission rate & $0.65/\text{wk}$ $[0.58 ; 0.73]$ \\
$a$ (determ.)& Amplitude of seasonality & $0.18$ $[0.17 ; 0.19]$ \\ 
$\delta$ (determ.)& Immunity waning rate & $1/(8.2\text{yr})$ $[1/12.0 ; 1/6.4]$\\
\midrule
$\beta_0$ (stoch.) & Mean transmission rate & $0.71/\text{wk}$ $[0.65 ; 0.77]$ \\
$a$ (stoch.) & Amplitude of seasonality & $0.18$ $[0.15 ; 0.20]$ \\
$\delta$ (stoch.) & Immunity waning rate & $1/(8.7\text{yr})$ $[1/12.0 ; 1/6.9]$\\
$\sigma$ (stoch.) & Level of stochasticity & $0.12$ $[0.11 ; 0.13]$\\
\midrule
Parameter & Description & Assumed value\\
\midrule
$\mu$ & Birth/death rate & $1/(80\text{yr})$ \\
$\gamma$ & Recovery rate & $1/(2.5\text{wk})$ \\

\bottomrule
\end{tabular}
\end{table}

\begin{figure}
\centering
\includegraphics[width=1.0\linewidth]{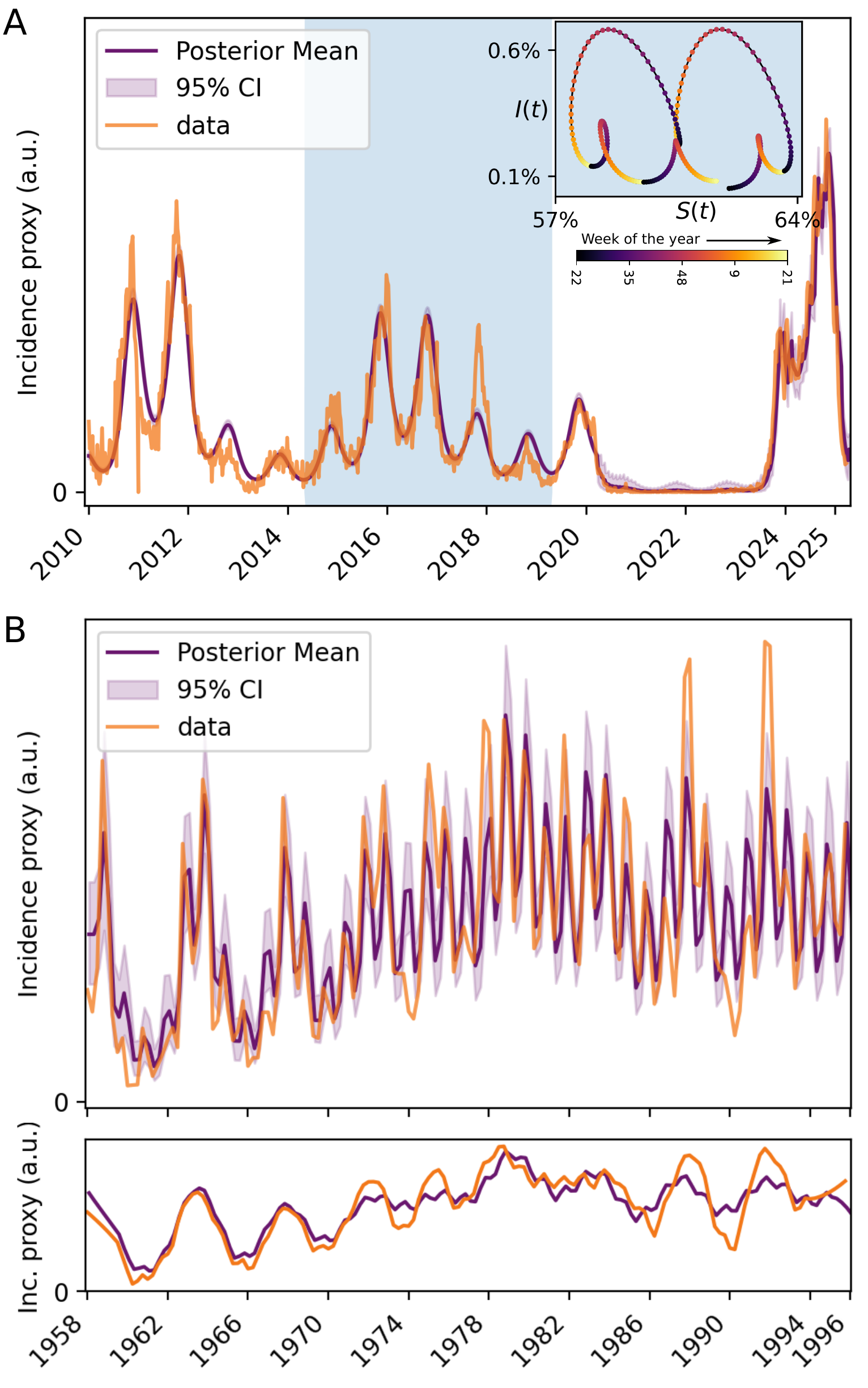}
\caption{\textbf{Model fits to \textit{Mycoplasma pneumoniae} incidence data.} A seasonal SIRS model is fitted to recent (2010--2025) and historical (1958--1995) \textit{M. pneumoniae} testing data from Denmark using Bayesian inference. \textbf{A)} The seasonal SIRS model explains observed \textit{M. pneumoniae} dynamics 2010--2025. The data are PCR positivity rates. The insert shows a full cycle (the interval highlighted in blue in the main plot) in the $(S,I)$ configuration space. Markers are one week apart and the color indicates time of year.  \textbf{B)} The same type of model captures key features of observed \textit{M. pneumoniae} dynamics 1958--1995. The fitted data are variance-stabilized numbers of samples which are simultaneously CA and MpCF-positive. The shallow bottom panel shows a smoothed version of the time series, to emphasize the multiannual peaks.
(Data source: Statens Serum Institut and \cite{lind1997seroepidemiological}).}
\label{fig:fit}
\end{figure}

\begin{figure}
\centering
\includegraphics[width=0.8\linewidth]{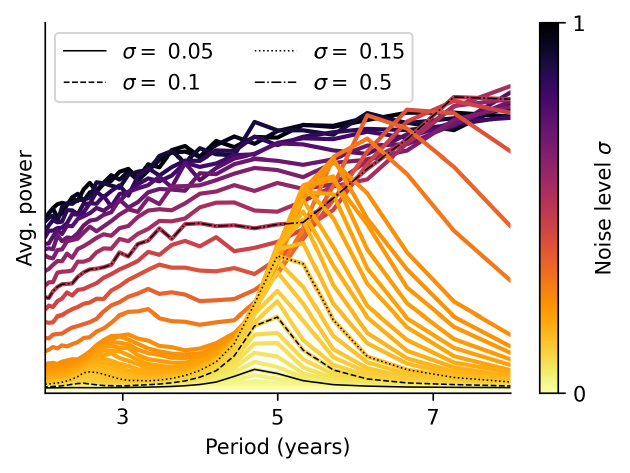}
\caption{\textbf{Moderate environmental noise stimulates multiannual cycles}, especially approximately 5-year cycles. Average Fourier spectrum across 40,000 realizations (for each noise level indicated in the legend) of the prevalence $I(t)$ of the seasonal SIRS model with parameters from the fit to 2010--2025 data from Denmark. The noise source is random fluctuations in the transmission rate $\beta(t)$, implemented as Gamma-distributed multiplicative noise with mean $1$ and variance $\sigma^2$ (see \textit{Materials and Methods} and \cite{he2010plug}). 80 years of each simulation time series were analyzed, and a 100 year ``transient'' was discarded. }
\label{fig:betafluc}
\end{figure}

\begin{figure}
\centering
\includegraphics[width=1.0\linewidth]{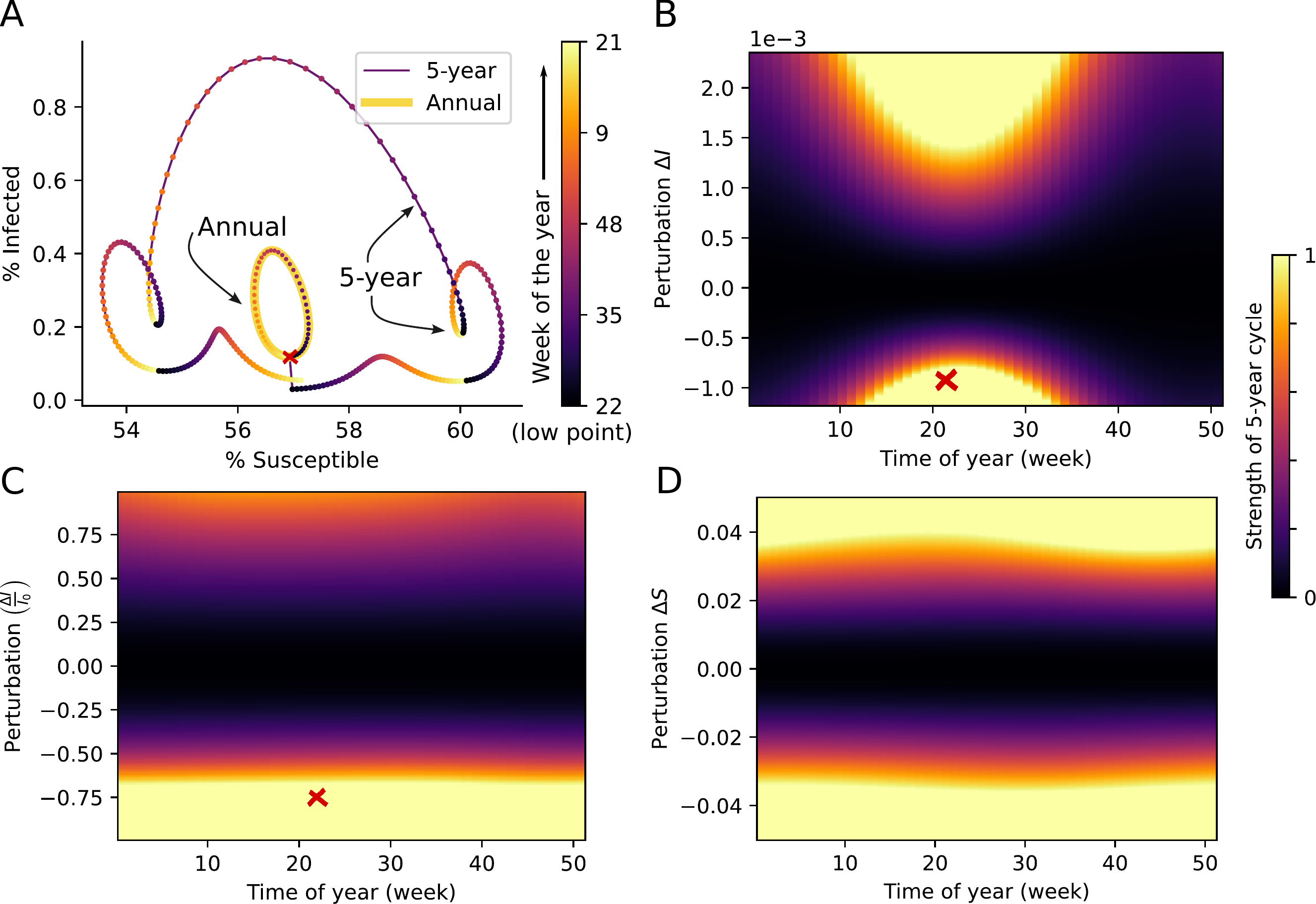}
\caption{\textbf{One-off exogeneous perturbations (``kicks'') transiently excite multiannual cycles.} \textbf{A)} System trajectory depicted in $(S,I)$ configuration space before and after a perturbation. The marker color denotes the time of year, and markers are one week apart. The approximately vertical line starting from the red cross shows an exogenous ``kick'' (a discrete change in prevalence, in this case) which causes the system to transition from the annual (thick yellow line) to 5-year cycle (thin purple line).
\textbf{B)} After a discrete change $\Delta I$ in the prevalence, the system may temporarily settle into a 5-year cycle, as in panel A. The plot shows how the size of the perturbation and the time of year affect the strength of the resulting cycle. The colorbar indicating the power of the 5-year cycle in the Fourier spectrum (relative to the highest peak) is shared between panels B, C, and D. Spectrum were based on the first 20 years post-perturbation. The red cross indicates the perturbation implemented in panel A. \textbf{C)} If the perturbation $\Delta I$ is proportional to the pre-perturbation prevalence $I_0$, the prominence of the 5-year cycle ceases to depend on time of year. The red cross indicates the perturbation implemented in panel A.  \textbf{D)} Effects of perturbations in $S(t)$, rather than the prevalence. }
\label{fig:kicktiming}
\end{figure}

\begin{figure}
\centering
\includegraphics[width=0.8\linewidth]{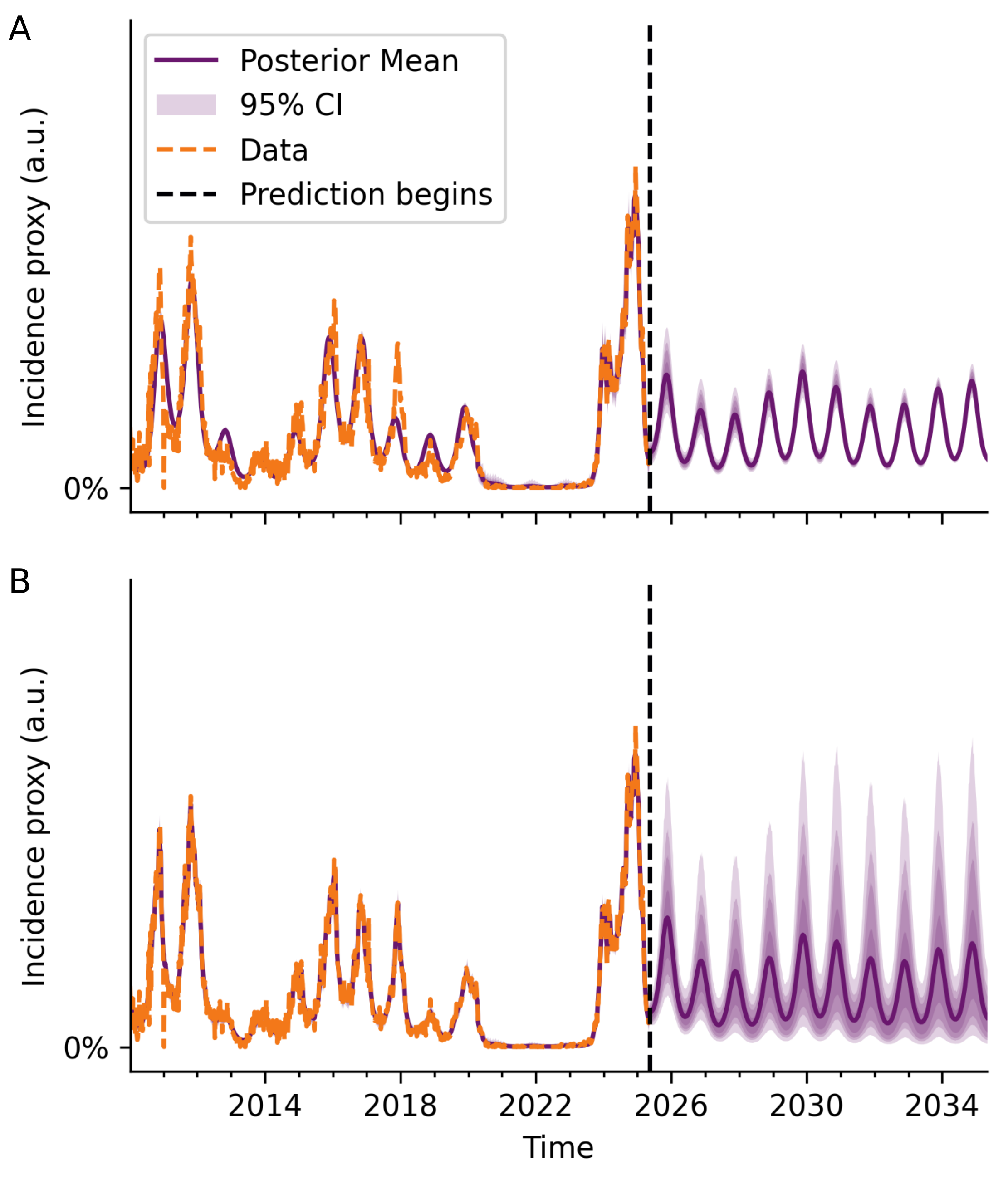}
\caption{\textbf{Prediction is hard.} 
\textbf{A)} Prediction using the deterministic seasonal SIRS model, parameters fitted to 2010--2025 data.
\textbf{B)} Prediction using the seasonal SIRS model with environmental stochasticity (noise in transmission, $\sigma = 0.12$, fitted value).
Noise
can significantly affect long- and medium-term predictions, but moderate to high \textit{M. pneumoniae} activity is predicted in the 2025/2026 season. The 95\% confidence interval is shown in light purple, and the 85\%, 75\%, and 50\% intervals are shown in successively darker shades.}
\label{fig:prediction}
\end{figure}

\matmethods{
\subsection*{The seasonal SIRS model}
The seasonally-forced SIRS model with vital dynamics and constant population size is given by
\begin{align}
	\od{S}{t} &= \mu - \beta(t) S(t) I(t) + \delta R(t) - \mu S(t), \label{eq:SIRS1}\\
	\od{I}{t} &= \beta(t) S(t) I(t) - (\gamma + \mu) I(t), \label{eq:SIRS2}\\
	\od{R}{t} &= \gamma I(t) - (\delta + \mu) R(t),\label{eq:SIRS3}
\end{align}
where the instantaneous transmission rate $\beta(t)$ is a non-negative and periodic function with a period of one year, i.e. $\beta(t + 1\text{yr}) = \beta(t)$. The remaining parameters are the birth/death rate $\mu$, recovery rate $\gamma$ and immunity waning rate $\delta$.
Noise in the transmission rate $\beta(t)$ is implemented as described in \cite{he2010plug}; when numerically integrating the above set of differential equations, we draw multiplicative noise from a Gamma distribution with shape $1/\sigma^2$ and scale $\sigma^2$. The parameter $\sigma$ is the noise level referred to in Fig. \ref{fig:betafluc}. 
\paragraph{Linearization of the autonomous system. } 
The analogous autonomous system to Eqs. \ref{eq:SIRS1}-\ref{eq:SIRS3} is obtained by substituting a constant transmission rate $\beta$ in place of $\beta(t)$. If perturbed away from equilibrium, $I(t)$ performs an underdamped oscillation while approaching the equilibrium configuration $(S^*,I^*,R^*)$, where:
\begin{align*}
    S^* = \frac{\gamma + \mu}{\beta}, ~~ I^* = \frac{(\beta-\mu-\gamma)(\delta + \mu)}{(\gamma + \delta + \mu)\beta}, ~~ R^* = \frac{(\beta-\gamma-\mu)\gamma}{(\gamma + \delta + \mu)\beta}.
\end{align*}
Linearizing the system around this equilibrium and applying a small perturbation, the decay rate and characteristic frequency are given by the real and imaginary parts of the eigenvalues of the Jacobian matrix:
\begin{align*}
    J = \left(
\begin{array}{ccc}
 \frac{\gamma  \delta -\beta  (\delta +\mu )}{\gamma +\delta +\mu } & -\gamma -\mu  & \delta  \\
 \frac{(\delta +\mu ) (\beta -\gamma -\mu )}{\gamma +\delta +\mu } & 0 & 0 \\
 0 & \gamma  & -\delta -\mu  \\
\end{array}
\right)
\end{align*}
The angular frequency of the intrinsic oscillation is then given by $\omega = \sqrt{\Omega_0^2 - (\Gamma/2)^2}$, where $\Omega_0^2=(\beta-\gamma-\mu)(\delta + \mu)$ and $\Gamma = \frac{(\beta + \delta) (\delta + \mu)}{\gamma + \delta + \mu}$. Since $\beta-\gamma$ is much greater than $\mu$ and $\delta$, it holds that $\omega \approx \Omega_0$, although we choose to keep $\Gamma$ in our calculations. The corresponding period is then given by
\begin{align}\label{eq:Tintrinsic}
    T = \frac{2\pi}{\sqrt{\Omega_0^2 - (\Gamma/2)^2}}. 
\end{align}
With the parameters fitted to the Danish 2010--2025 time series, the eigenvalues are $\lambda_1 = -2.4\times 10^{-4}\text{wk}^{-1}$ and $\lambda_\pm = (-0.0020 \pm 0.024 i)\text{wk}^{-1}$. Since all real parts are strictly negative, the equilibrium is asymptotically stable.
\subsection*{Bayesian fitting of a deterministic model to 2010--2025 data}
We fit the seasonal SIRS model as described above using the probabilistic programming language Stan \cite{stan,cmdstanpy}, with a sinusoidal seasonally forced transmission rate 
$\beta(t)= \beta_0 \xi(t)  (1 + a \sin(2 \pi t/\text{year} + \phi))$, to positivity rates given in the 2010--2025 data set. $\xi(t)$ is the NPI multiplier which is fixed to 1 until March 2020 and otherwise estimated.
The priors used were:
\begin{itemize}
    \item $\beta_0 \sim N[0; 5/T]$ where $N[x; y]$ indicates a Gaussian (normal) distribution with mean $x$ and standard deviation $y$, while $T=1/\gamma$ is the recovery time (2.5wk).
    \item $\text{NPI multipliers } \xi(t) \sim N[0.8; 0.3]$
    \item $\log(\rho) \sim N[-2; 0.5]$
    \item $S(t=0) \sim N[0.3; 0.3]$
    \item $\log(x_I) \sim N[-4; 2]$ such that $I(t=0) = x_I S(t=0)$ (the stick-breaking method, ensuring $S(t=0) + I(t=0) \leq 1$)
    \item $\phi \sim U[0; 2\pi]$ where $U[x;y]$ indicates the uniform distribution with support on the interval $[x, y]$
    \item $a \sim N[0; 0.10]$
    \item $\delta \sim N[1/(20\text{yr}); 1/(10\text{yr})]$
    \item $\rho x_\text{data} \sim N[x_\text{model}; \sigma_\text{obs}]$
    \item $\sigma_\text{obs} \sim N[0; 0.1]$
\end{itemize}
Here $x_\text{data}$ and $x_\text{model}$ refer to the measured incidence proxy (positivity rate, arbitrarily rescaled) and the modeled incidence, respectively. 
Note that for the parameters $\beta_0$, $\rho$, $S(t=0)$, $a$, $\delta$, $x_\text{data}$, $\sigma_\text{obs}$ and the NPI multipliers, only the part of the normal distribution with positive support is used, since the parameters in question are necessarily non-negative. The fit to the 1958--1995 MpCF-based data follows similar principles and is detailed in the appendix. For the purposes of fitting, the model (Eqs. \ref{eq:SIRS1}-\ref{eq:SIRS3}) is implemented using the efficient discretization scheme described in \cite{park2024predicting}, with a time-step of $0.125\text{wk}$. This discretization was used for Figs. \ref{fig:fit}, \ref{fig:betafluc}, \ref{fig:kicktiming}, \ref{fig:prediction}, S4, S5, S7, S8 and S10,
whereas the calculations of Figs. S1 %
and S9 
were performed using standard Euler integration of Eqs. \ref{eq:SIRS1}-\ref{eq:SIRS3}.
Convergence was assessed based on the absence of diagnostic warnings from CmdStanPy \cite{cmdstanpy}, including the absence of divergent chains, no exceedances of tree-depth,  sufficient E-BFMI ($>0.3$), and satisfactory $\hat{R}$ ($<1.05$) and effective sample size (bulk $\text{ESS}>100\times \text{\#chains}$) for all parameters.

\paragraph*{Bayesian fitting of a stochastic model}
We fit a seasonal SIRS model with environmental stochasticity in the form of Gamma-distributed multiplicative noise in the transmission rate, as used in Fig. \ref{fig:prediction}. The model set-up, using Stan \cite{stan}, closely follows that of the deterministic model. A new fitted parameter, the noise level $\sigma$, is introduced which controls the magnitude of noise in $\beta(t)$, with prior $\sigma \sim N(0.1, 0.02)$ (positive-support part of normal distribution only). The individual perturbations are then implemented by letting $\beta(t) \to \eta(t) \beta(t)$ where $\eta \sim \Gamma[1/\sigma^2, \sigma^2]$ with the notation $\Gamma[\alpha,\theta]$ referring to a Gamma distribution with shape $\alpha$ and scale $\theta$. For computational efficiency, we only draw a new value of $\eta(t)$ every other week.

\subsection*{Doob-Gillespie simulations}
The stochastic simulations of Fig. \ref{fig:demographic} are based on the following system of transition rates:
\begin{align}
	r[\emptyset \to S]  &= \mu N, \nonumber \\
	r[S \to \emptyset] 	&= \mu S, \nonumber \\
	r[I \to \emptyset]	 &= \mu I, \nonumber \\
	r[R \to \emptyset]  &= \mu R, \nonumber \\
	r[S \to I]			&= 
    \begin{cases}
        \beta(t) S I / N & \text{if } N > 0,\\
        0  & \text{otherwise}, 
    \end{cases} \nonumber \\
	r[I \to R]					&= \gamma I, \nonumber \\
	r[R \to S]    &= \delta R, \label{eq:gilrates}
\end{align}
with the time-dependent transmission rate given by $\beta(t)= \beta_0  (1 + a \sin(2 \pi t/\text{year} + \phi))$. Note that here, $S$, $I$, and $R$ denote the populations of susceptible, infected, and recovered individuals (as integers), not the population fractions.
Denoting the above rates by $r_i$, $i \in \{1,\dots,7\}$, the Doob-Gillespie algorithm \cite{gillespie1977exact} thus prescribes iteratively propagating the state of the system by drawing a uniformly distributed random number $u_1 \in [0,1]$, determining the time to next event $\tau = -(\sum_i r_i)^{-1} \log(u_1)$ and the nature of the event by drawing another random uniform number $u_2 \in [0,1]$ and determining the smallest $j$ for which $\sum_{i=1}^j r_i > u_2 \sum_{i=1}^7 r_i$. Event (or ``reaction'') $j$ is then performed and the time is incremented by $\tau$. The procedure is then reiterated until the desired final time is reached. The metapopulation simulations of Supplementary Fig. S11 %
use the same algorithm, but with modified rates. Details are given in the supplement.
}

\showmatmethods{} %

\acknow{We thank Statens Serum Institut (SSI) for providing Danish surveillance data. We would also like to thank Mercedes Pascual for highly valuable input. BFN acknowledges financial support from the Carlsberg Foundation (grants CF23-0173 and CF24-1337). The PandemiX Center is funded by the Danish National Research Foundation (grant DNRF170). SWP is a Peter and Carmen Lucia Buck Foundation Awardee of the Life Sciences Research Foundation. MHJ is supported by the Novo Nordisk Foundation, grants NNF20OC0064978 and NNF24OC0089788. BTG would like to acknowledge support from Princeton Catalysis Initiative and Princeton Precision Health.}

\showacknow{} 

\paragraph*{CODE AND DATA.}
All code and data used in this manuscript may be accessed at \url{https://github.com/BjarkeFN/MycoplasmaDynamics} . 

\bibliography{biblio}
\newpage
\FloatBarrier

\begin{figure}
\centering
\includegraphics[width=0.8\linewidth]{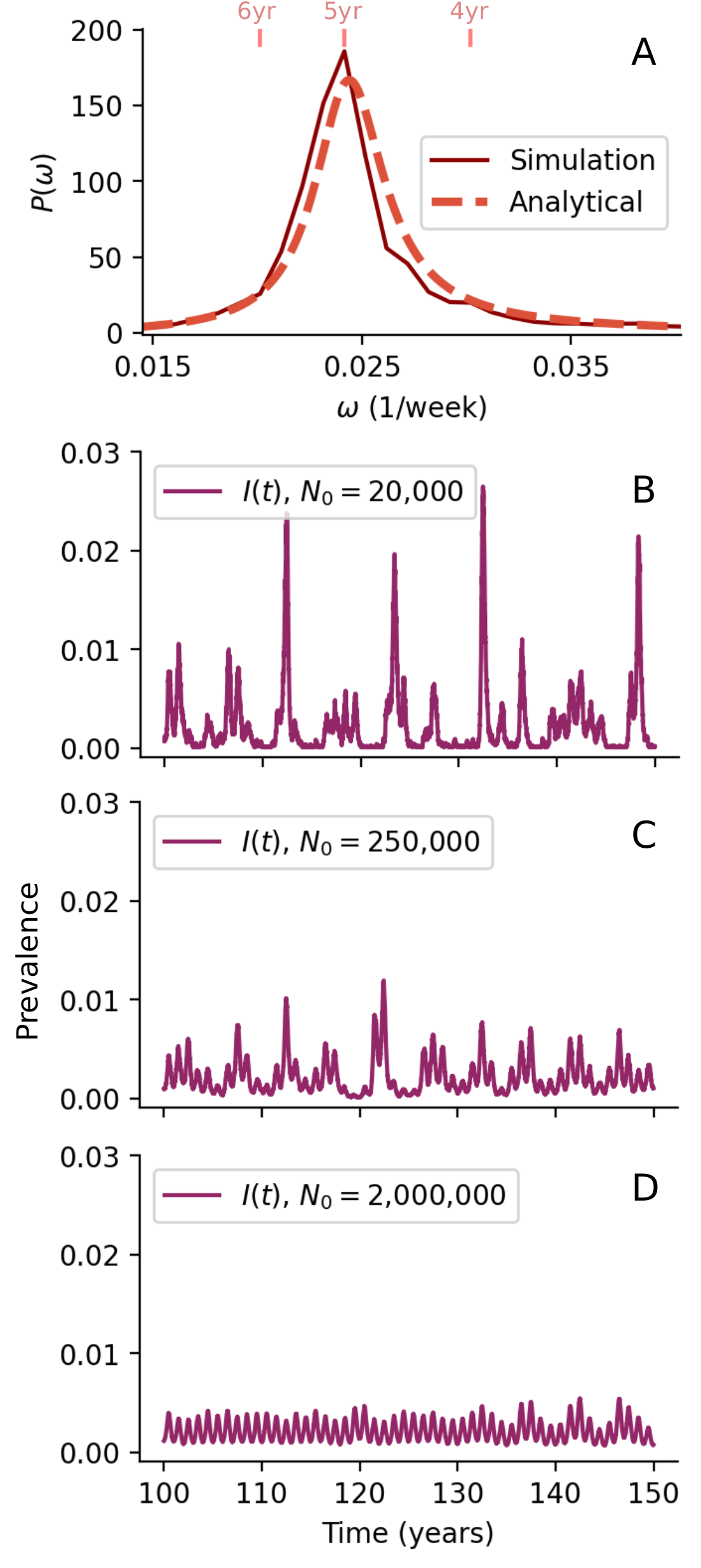}
\caption{\textbf{Demographic stochasticity excites multi-year cycles in smaller populations.} \textbf{A)} Theoretical power spectral density (PSD) of multiannual fluctuations excited by demographic noise (without seasonality), and avg. power spectrum of 50 stochastic simulations (with seasonality). See Supplementary Fig. S3 %
for plots of the dependence of the theoretical PSD on individual parameters. \textbf{B)} Stochastic simulation of a seasonally forced SIRS model parametrized with \textit{M. pneumoniae} parameters as determined by Bayesian fit to 2010--2025 data. In small populations (initial population size $N_0 = 20,000$), demographic stochasticity leads to noisy multi-year cycles. \textbf{C)} In intermediate populations ($N_0=250,000$), multi-year cycles are prominent and fairly regular. \textbf{D)} In larger populations ($N_0=2\times 10^6$), annual dynamics dominates as the impact of demographic stochasticity is smaller.}
\label{fig:demographic}
\end{figure}

\FloatBarrier


\section*{Supplementary material}

# SI Appendix

**Quantifying chaos: Lyapunov exponents**   In Fig. S1A, the instantaneous Lyapunov exponents $\lambda_i(t)$ of the system are given. In Fig. S1B, we present the mean Lyapunov exponents during the first $T$ years, i.e. $\frac{1}{T}\int_0^T \lambda(t)dt$. Local Lyapunov exponents were computed using the QR method described in (55).

**Bayesian fitting of 1958-1995 MpCF data**   As for the 2010-2025 data, we fit a seasonal SIRS model using the probabilistic programming language Stan (67, 68), with a sinusoidal seasonally forced transmission rate $\beta(t) = \beta_0(1 + a\sin(2\pi t/\text{year} + \phi))$. We fit the modeled incidence to the variance-stabilized (square root-transformed) number of samples simultaneously positive for CA and MpCF (test data from (2)). There are no NPI multipliers in this fit, but to account for slowly changing contact rates and ascertainment patterns over the years, we included multipliers for the transmission rate $\beta(t)$ and the $\rho$, the constant of proportionality between measured and simulated incidence. These multipliers had relatively narrow priors centered around unity and were piecewise constant for each four years of the simulation. The priors used for the 1958-1995 fit were mostly similar to those for the 2010-2025 fit, but due to the superior data quality of the 2010-2025 data, we constrained some variables using results from that fit:

- $\beta_0 \sim N[0; 5/T]$ where $N[x; y]$ indicates a Gaussian (normal) distribution with mean $x$ and standard deviation $y$. However, the variable $\beta_0$ was itself constrained to the interval $[1.3/T; 2.2/T]$.
- $\log(\rho) \sim N[-2; 0.5]$
- $S(t=0) \sim N[0.3; 0.3]$
- $\log(x_I) \sim N[-4; 2]$ such that $I(t=0) = x_I S(t=0)$ (the stick-breaking method, ensuring $S(t=0) + I(t=0) \leq 1$)
- $\phi \sim U[0; 2\pi]$ where $U[x; y]$ indicates the uniform distribution with support on the interval $[x, y]$
- $a \sim N[0; 0.10]$
- $\beta(t)$ multipliers $\sim N[1; 0.05]$
- $\rho$ multipliers $\sim N[1; 0.2]$
- $\rho x_\text{data} = N[x_\text{model}; \sigma_\text{obs}]$
- $\sigma_\text{obs} \sim N[0; 0.01]$

Here $x_\text{data}$ and $x_\text{model}$ refer to the measured incidence proxy and the modeled incidence, respectively. $\rho$ is a constant of proportionality between the two. Note that for the parameters $\beta_0$, $\rho$, $S(t=0)$, $a$, $x_\text{data}$, $\sigma_\text{obs}$ and the $\beta(t)$ and $\rho$ multipliers, only the part of the normal distribution with positive support is used, since the parameters in question are necessarily non-negative. The immunity rate of waning $\delta$ was fixed at the value obtained in the 2010-2025 plot.

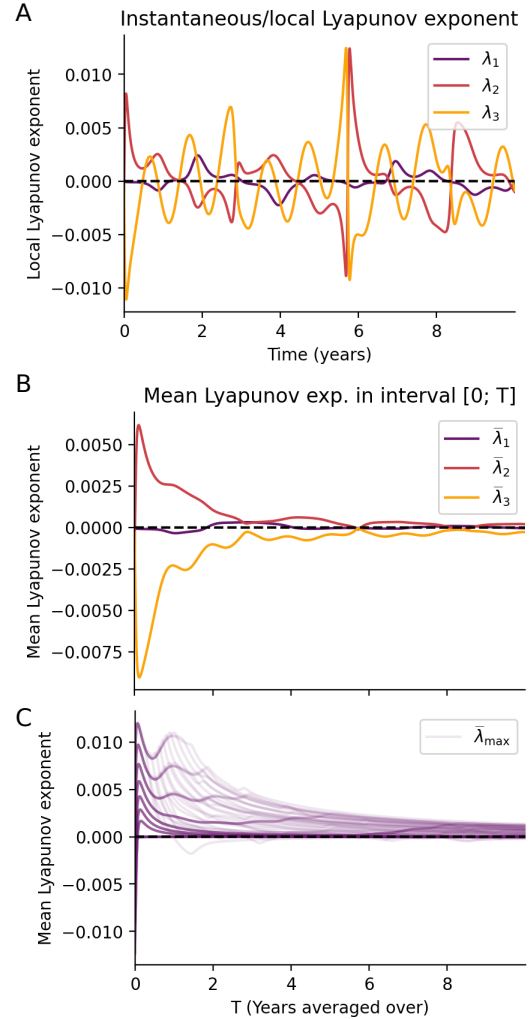

**Fig. S1. Lyapunov exponents show transient chaos A)** Local Lyapunov exponents $\lambda_i(t)$. The initial condition (at $t=0$) corresponds to the initial state (January 2010) of the 2010-2025 fit. **B)** Cumulative mean local Lyapunov exponents during the first $T$ years, i.e. $\frac{1}{T}\int_0^T \lambda(t)dt$. **C)** Same as panel B, but for a wide range of initial conditions, and only the largest Lyapunov exponent is shown.

**Table S1. Estimated and assumed parameters from fit to 1958-1995** *Mycoplasma pneumoniae* **surveillance data from Denmark.**

| Parameter | Description | Estimated value [95% HPD] |
|---|---|---|
| $\beta_0$ | Mean transmission rate | $0.67/\text{wk}$ [0.63; 0.72] |
| $a$ | Amplitude of seasonality | $0.124$ [0.097; 0.152] |

| Parameter | Description | Assumed value |
|---|---|---|
| $\mu$ | Birth/death rate | $1/(75\text{yr})$ |
| $\gamma$ | Recovery rate | $1/(2.5\text{wk})$ |
| $\delta$ | Rate of waning of immunity | $1/(11.2\text{yr})$ (from 2025 fit) |

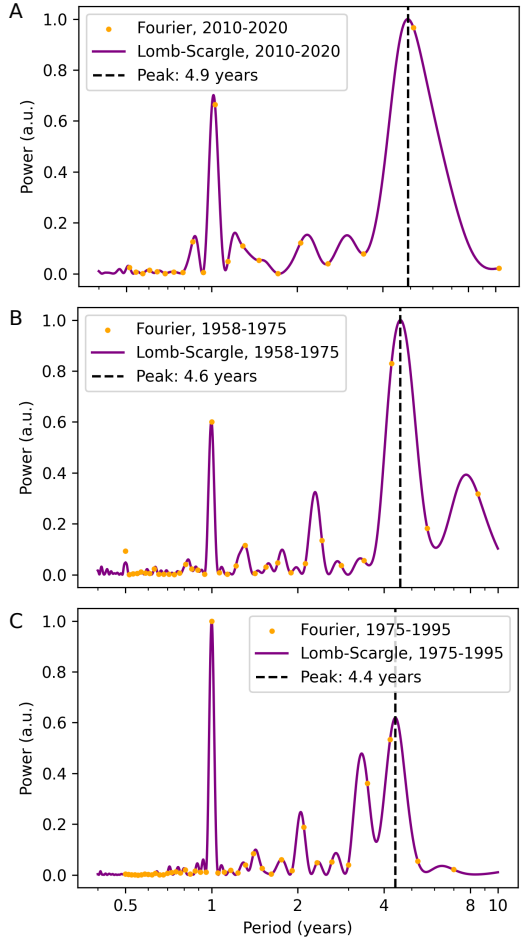

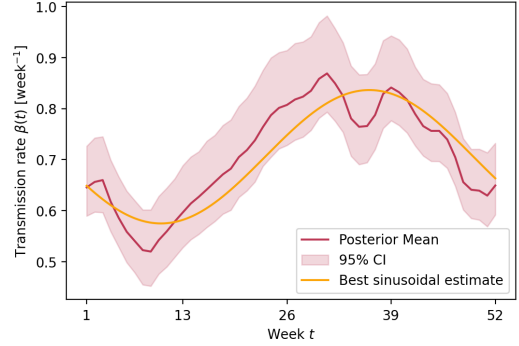

**Fig. S4. Seasonally varying transmission rate.** The dark red curve is the transmission rate obtained when fitting a seasonal profile. The orange curve is the transmission rate when instead imposing that $\beta(t) = \beta_0(1 + a\sin(2\pi t/\text{year} + \phi))$ and fitting the parameters $\beta_0$, $a$, and $\phi$.

**Fig. S2. Fourier spectra and Lomb-Scargle periodograms for Danish *M. pneumoniae* incidence time series. A)** Spectra for the period January 2010 - March 2020 (prior to COVID-19 public health interventions). **B)** Spectra for the period February 1958 - January 1975. **C)** Spectra for the period January 1975 - November 1995.

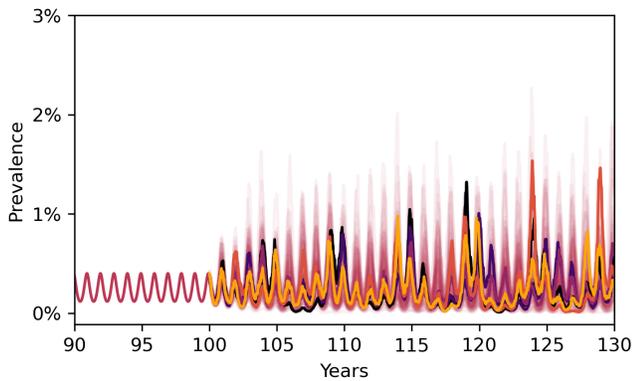

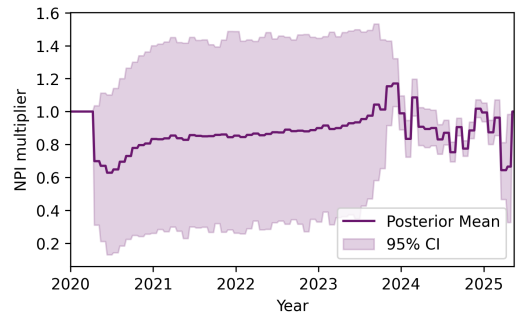

**Fig. S5. NPI multipliers.** Estimated transmission rate modulating factor. Until March 2020, this multiplier was fixed at $1.0$.

**Fig. S3. Noise in transmission excites multiannual cycles starting from the annual attractor.** After allowing the system to settle on the annual attractor, noise in the transmission rate $\beta(t)$ is "turned on" (noise level $\sigma = 0.15$) at $t = 100$ yr. 100 realizations are shown, with five of them highlighted to clearly show the multiannual cycles which develop.

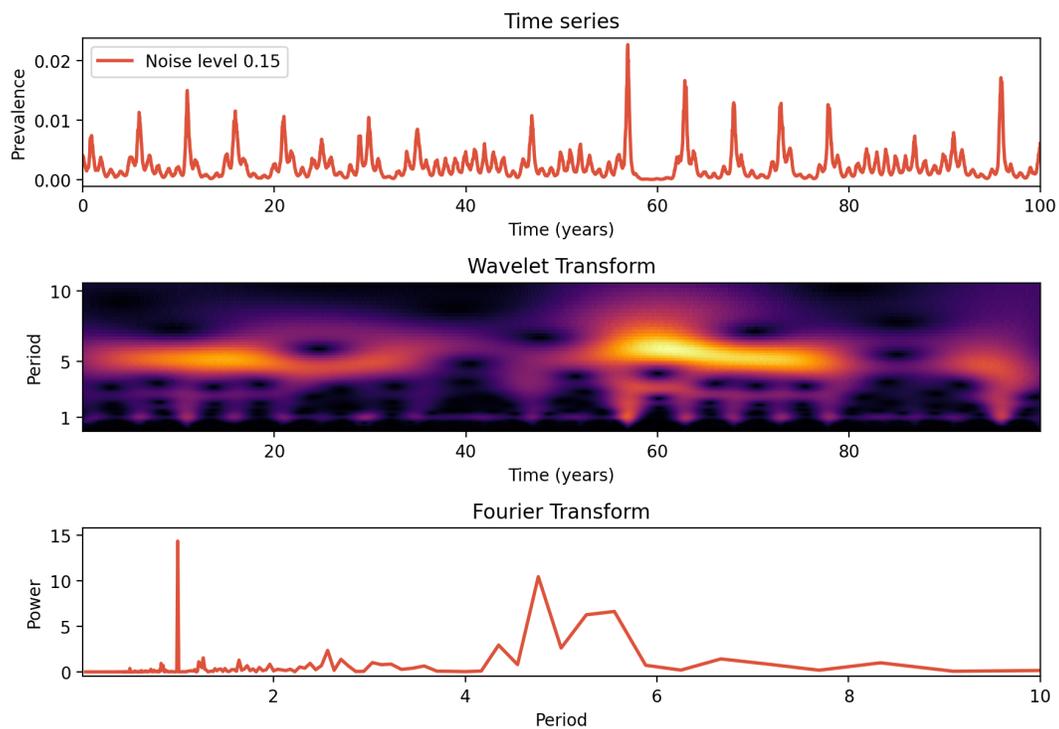

**Fig. S6. The seasonal SIRS model with (environmental) noise exhibits mode-hopping. A)** 100-year simulated time-series with noise level $\sigma = 0.15$. **B)** Continuous wavelet transform of the signal from panel A. **C)** Fourier transform of the entire 100-year time series. The wavelet transform confirms that the system transitions in and out of an approximately 5-year cycle, sometimes displaying primarily annual dynamics.


\end{document}